\documentclass{jpsj2}
\usepackage[nomarkers]{endfloat}

\newcommand{\Tss}[3]{ \ensuremath{#1 _{\mathrm{#2}} ^{\mathrm{#3}} }}
\newcommand{\Tsub}[2]{\ensuremath{#1_{\mathrm{#2}}}}
\newcommand{\Tsup}[2]{\ensuremath{#1^{\mathrm{#2}}}}
\newcommand{\TT}[1]{\Tsub{T}{#1}}
\newcommand{\TTp}[1]{\Tsup{T}{#1}}
\newcommand{\Tk}[1]{\Tsub{\kappa}{#1}}
\newcommand{\TC}[1]{\Tsub{C}{#1}}
\newcommand{\Tv}[1]{\Tsub{v}{#1}}
\newcommand{\Tl}[1]{\Tsub{l}{#1}}
\newcommand{\Tt}[1]{\Tsub{\tau}{#1}}
\newcommand{\Tkai}[1]{\Tsub{\chi}{#1}}
\newcommand{\TJ}[2]{\Tss{J}{#1}{#2}}
\newcommand{\TTst}{\TTp{*}}
\newcommand{\TDebye}{\Tsub{\Theta}{D}}
\newcommand{\TkB}{\Tsub{k}{B}}
\newcommand{\TSCO}[3]{Sr$_{#1}$Cu$_{#2}$O$_{#3}$}

\newcommand{\Tion}[2]{\ensuremath{\mathrm{#1}^{\mathrm{#2}}}}
\newcommand{\TK}{\ensuremath{ \mathrm{K} }}
\newcommand{\Tmm}{\ensuremath{ \mathrm{mm} }}
\newcommand{\TkOhm}{\ensuremath{ \mathrm{k \Omega} }}
\newcommand{\TdegC}{\ensuremath{\mathrm{\char'27 \kern-.2em \hbox{C}}}}
\newcommand{\TmSR}{\ensuremath{ \mu \mathrm{SR} }}
\newcommand{\Teqref}[1]{(\ref{#1})}
\newcommand{\TTabSpace}[1]{\hspace{.5em}{#1}\hspace{.5em}}
\newcommand{\TFigScale}{1}
\newcommand{\TFigScaleL}{1}
\begin{document}

\title{Evidence for Ballistic Thermal Conduction in the One-Dimensional $S=1/2$ Heisenberg Antiferromagnetic Spin System Sr$_2$CuO$_3$}

\author{
	Takayuki \textsc{Kawamata}$^{1, 2}$
	\thanks{
		E-mail address: tkawamata@riken.jp, 
		Present address: Advanced Meson Science Laboratory, RIKEN (The Institute of Physical and Chemical Research), 2-1 Hirosawa, Wako 351-0198
	}, 
	Nobuo \textsc{Takahashi}$^{1}$, 
	Tadashi \textsc{Adachi}$^{1, 2}$, 
	Takashi \textsc{Noji}$^{1, 2}$, 
	Kazutaka \textsc{Kudo}$^{3}$, 
	Norio \textsc{Kobayashi}$^{3}$ 
	and	
	Yoji \textsc{Koike}$^{1, 2}$
}

\inst{
	$^{1}$Department of Applied Physics, Tohoku University, 6-6-05 Aoba, Aramaki, Aoba-ku, Sendai 980-8579
	
	$^{2}$CREST, Japan Science and Technology Corporation (JST)
	
	$^{3}$Institute for Materials Research, Tohoku University, 2-1-1 Katahira, Aoba-ku, Sendai 980-8577
}

\abst{
We have measured the thermal conductivity of the one-dimensional (1D) $S=1/2$ Heisenberg antiferromagnetic spin system of Sr$_2$Cu$_{1-x}$Pd$_x$O$_3$ single crystals including nonmagnetic impurities of Pd$^{2+}$. 
It has been found that the mean free path of spinons along the 1D spin chain at low temperatures is very close to the average length of finite spin chains between spin defects estimated from the magnetic susceptibility measurements. 
This proves that the thermal conduction due to spinons at low temperatures in Sr$_2$CuO$_3$ is ballistic as theoretically expected [ Zotos \textit{ et al.}, Phys. Rev. Lett. \textbf{55} (1997) 11029].
}

\kword{ballistic thermal conduction, thermal conductivity, one-dimensional quantum spin system, Sr$_2$CuO$_3$}

\maketitle

\section{Introduction}
Recently, thermal conductivity in low-dimensional quantum spin systems with the spin quantum number $S=1/2$ has attracted interest, because the thermal conductivity due to spin excitations, \Tk{spin}, has been found to be high in various materials, 
such as the 2-leg spin ladder system \TSCO{14}{24}{41} \cite{Kudo:JLTP117:1999:1689,Kudo:JPSJ70:2001:437,Sologubenko:PRL84:2000:2714,Hess:PRB64:2001:184305,Hess:PRL93:2004:027005}, 
one-dimensional (1D) antiferromagnetic (AF) spin systems Sr$_2$CuO$_3$ \cite{Sologubenko:PRB62:2000:R6108,Sologubenko:PRB64:2001:054412}, 
\TSCO{}{}{2} \cite{Sologubenko:PRB64:2001:054412}, 
Bechgaard salts \cite{Lorenz:N418:2002:614} 
and BaCu$_2$Si$_2$O$_7$ \cite{Sologubenko:EL62:2003:540}. 
It has been found that one of essential factors of high \Tk{spin} is a large bandwidth of the spin excitations bringing on a high velocity of the spin excitations and that the AF correlation between the nearest neighboring spins is more suitable for high \Tk{spin} than the ferromagnetic one \cite{Kudo:JMMM272:2004:94}.
However, the mechanism of \Tk{spin} in low-dimensional quantum spin systems has not been understood fully.

Accordingly to the theoretical study using the Kubo formula \cite{Kubo:JPSJ12:1957:570}, it has been predicted that the thermal conduction due to spin excitations is ballistic at finite temperatures in 1D spin systems with $S = 1/2$ described by integrable Hamiltonian's, because the heat flow is a conserved quantity \cite{Castella:PRL74:1995:972,Saito:PRE54:1996:2404,Zotos:PRB55:1997:11029,Zotos:PRL82:1999:1764,Klumper:JPA35:2002:2173}. 
That is, in these systems, the thermal conductivity due to spin excitations, namely, due to spinons possessing $S = 1/2$, \Tk{spinon}, is expected to be very high, because the mean free path of spinons, \Tl{spinon}, is infinite in the ideal case. 
In spin systems described by non-integrable Hamiltonian's, on the other hand, it has been predicted that the thermal conduction due to spins is diffusive and that \Tl{spinon} is comparable to the distance between the nearest neighboring spins in the high temperature limit.

As for Sr$_2$CuO$_3$, it has been found from magnetic susceptibility \cite{Ami:PRB51:1995:5994,Motoyama:PRL76:1996:3212}, specific heat \cite{Motoyama:PRL76:1996:3212} and midinfrared optical absorption measurements \cite{Suzuura:PRL76:1996:2579} that the intrachain exchange interaction, $\TJ{}{}$, is as large as more than $2000~\TK$.
On the other hand, the interchain exchange interaction, $\TJ{}{\prime}$, is as small as $\sim 10^{-4} \TJ{}{}$, according to the estimate from the AF transition temperature, \TT{N}, $\sim 5.4~\TK$ \cite{Keren:PRB48:1993:12926,Kojima:PRL78:1997:1787} using the simple relation, $\TT{N} \sim \sqrt[\leftroot{0} \uproot{2} 3]{\TJ{}{} \times \TJ{}{\prime 2}}$. 
Therefore, Sr$_2$CuO$_3$ is regarded as an almost ideal 1D $S=1/2$ Heisenberg AF spin system described by the integrable Hamiltonian, 
\begin{equation}
	H = J \sum _{i}  \boldsymbol{S} _{i} \cdot \boldsymbol{S} _{i+1}. 
	\label{equ:Hami}
\end{equation}
Actually, a high \Tk{spinon} has been observed in Sr$_2$CuO$_3$, as mentioned above \cite{Sologubenko:PRB62:2000:R6108,Sologubenko:PRB64:2001:054412}. 
The contribution of \Tk{spinon} to the thermal conductivity in Sr$_2$CuO$_3$ is characterized by a clear shoulder around $70~\TK$ in the temperature dependence of the thermal conductivity along the direction parallel to the spin chain only.
The existence of the ballistic thermal conduction due to spinons has been insisted from the analysis of \Tl{spinon}.
Besides, it has been reported from the NMR measurement of Sr$_2$CuO$_3$ that the spin-diffusion constant related to \Tk{spinon} is very large \cite{Takigawa:PRL76:1996:4612,Takigawa:PRB56:1997:13681}, though there is a report that the spin transport is diffusive at finite temperatures \cite{Thurber:PRL87:2001:247202}. 
Moreover, a similar large spin-diffusion constant has also been obtained from the NMR experiment in the 1D $S=1/2$ AF spin chain ststem $\alpha$-VO(PO$_3$)$_2$ \cite{Kikuchi:JPSJ70:2001:2765}. 

According to the discussion by Sologubenko \textit{et al.} \cite{Sologubenko:PRB62:2000:R6108,Sologubenko:PRB64:2001:054412} their insistence on the existence of the ballistic thermal conduction due to spinons in Sr$_2$CuO$_3$ is based on the result that the characteristic temperature of the spinon scattering is the order of the Debye temperature, {\TDebye}. 
It may follow the major scattering is spinon-phonon scattering rather than spinon-spinon scattering, but it seems too rough to insist that the thermal conduction due to spinons is ballistic because of the absence of the spinon-spinon scattering. 
Therefore it is not clear whether the thermal conduction due to spinons is ballistic or not. 
In this paper, in order to confirm the ballistic nature of the thermal conduction due to spinons in Sr$_2$CuO$_3$, we have grown Sr$_2$Cu$_{1-x}$Pd$_x$O$_3$ single crystals with $x = 0$, $0.004$, $0.010$ including nonmagnetic impurities of \Tion{Pd}{2+}, in which the average length of finite spin chains between spin defects, \Tsub{L}{imp}, is expected to decrease with increasing $x$. 
We have measured the thermal conductivity along the $b$-axis parallel to the 1D spin chain, \Tk{b}, and along the $a$-axis perpendicular to the 1D spin chain, \Tk{a}. 
We have measured the specific heat to estimate the value of {\TDebye} and the specific heat of spinons, \TC{spinon}. 
We have also measured the magnetic susceptibility to estimate the amount of spin defects. 
Then, we have compared the value of \Tl{spinon} estimated from \Tk{spinon} and \TC{spinon} with \Tsub{L}{imp} estimated from the magnetic susceptibility measurements.
As a result, it has been found that the value of \Tl{spinon} at low temperatures is very close to the value of \Tsub{L}{imp}, meaning that spinons are moving along the spin chain between spin defects without being scattered at low temperatures. 
This proves that the thermal conduction due to spinons at low temperatures in Sr$_2$CuO$_3$ is ballistic.
The preliminary results have already been reported in our previous paper \cite{Takahashi:ACP850:2006:1265}.


\section{Experimental}
Single crystals of Sr$_2$Cu$_{1-x}$Pd$_x$O$_3$ with $x = 0$, $0.004$, $0.010$ were grown by the Traveling-Solvent Floating-Zone (TSFZ) method. 
In order to prepare the feed rod for the TSFZ growth, first, we prepared polycrystalline powder of Sr$_2$Cu$_{1-x}$Pd$_x$O$_3$ by the solid-state reaction method.
The prescribed amount of SrCO$_3$, CuO and PdO powders with 99.9~\% purity was mixed, ground, and prefired at $800 \TdegC$ in air for 24 h. 
After pulverization, the prefired powder was mixed and sintered at $1030 \TdegC$ in air for 24 h with several times of intermediate grinding. 
After thorough grinding, the powder was isostatically cold pressed at 2.6 kbar into a rod of 7 mm in diameter and $\sim 120$ mm in length. 
Then, the rod was prefired at $800 \TdegC$ in air for 12 h and sintered at $1040 \TdegC$ in air for 24 h.
As a result, a tightly and densely sintered feed rod was prepared.
As Sr$_2$Cu$_{1-x}$Pd$_x$O$_3$ melts incongruently \cite{Nevriva:PC235:1994:325}, solvent disks with the composition of Sr : Cu : Pd $ = 45 : 55(1-x) : 55 x$ in the molar ratio were prepared in a similar way. The sintering was performed at $800 \TdegC$ in air for 12 h.
The TSFZ growth was carried out in flowing O$_2$ gas of 1 bar in an infrared heating furnace.
The rotation speed of the upper and lower shafts was 15 rpm in the opposite direction.
The zone traveling was 1.0 mm/h.
The grown crystals were annealed at $870 \TdegC$ for $72$ h in flowing Ar gas of 1 bar in order to remove excess oxygen. 
The crystals were characterized using x-ray back-Laue photography and were confirmed to be a single phase by powder x-ray diffraction.
The chemical composition of the crystal was determined by inductively coupled plasma optical emission spectrometry (ICP-OES) and comfirmed to be the same as the nominal composition. 

Thermal conductivity measurements were carried out by the conventional steady-state method. 
One side of a rectangular single-crystal, whose typical dimensions were $ 3.5 \times 0.8 \times 0.8~\Tmm^3 $, was anchored on the copper heat sink with indium solder. 
A chip-resistance of $1~\TkOhm$ (Alpha Electronics Corp. MP1K000) was attached as a heater to the opposite side of the single crystal with GE7031 varnish. 
The temperature difference across the crystal ($0.02$ -- $1.0 ~\TK$) was measured with two Cernox thermometers (Lake Shore Cryotronics, Inc. CX-1050-SD). 
The specific heat was measured by the thermal relaxation technique using a PPMS (Quantum Design, Model PPMS).
The magnetic susceptibility was measured using a SQUID magnetometer (Quantum Design, Model MPMS). 

\section{Results and Discussion}
\subsection{Magnetic susceptibility}
\begin{figure}[tbp]
	\begin{center}
		\includegraphics[scale=\TFigScale]{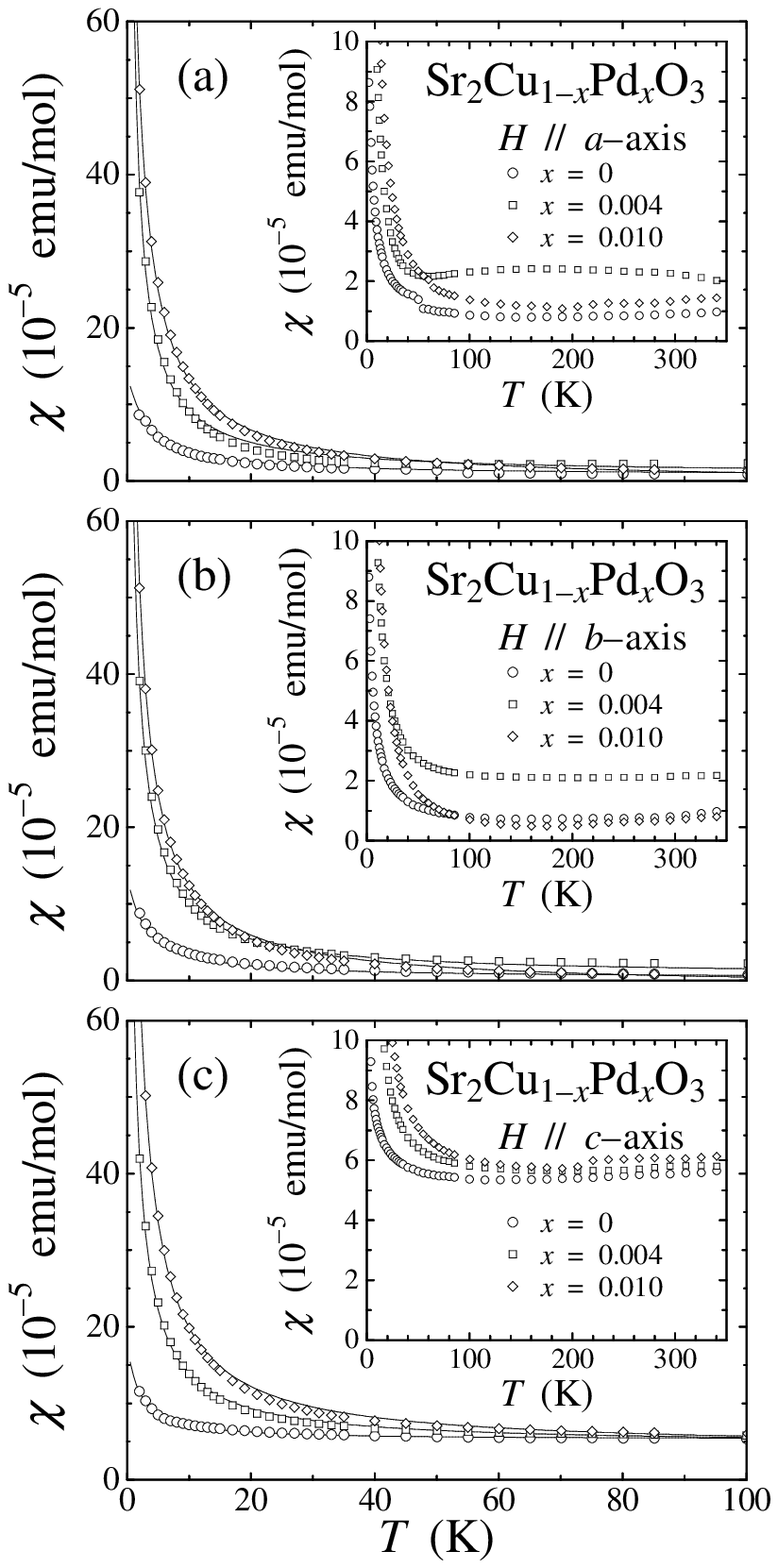}
		\caption{Temperature dependence of the magnetic susceptibility, \Tkai{}, of Sr$_2$Cu$_{1-x}$Pd$_x$O$_3$ with $x = 0$, $0.004$, $0.010$ in a magnetic field of 1 T parallel to (a) the $a$-axis, (b) $b$-axis and (c) $c$-axis. 
		Solid lines indicate the best-fit results using Eqs. \Teqref{equ:KaiAll} -- \Teqref{equ:KaiSpin}. }
		\label{fig:sup}
	\end{center}
\end{figure}
Figure \ref{fig:sup} shows the temperature dependence of the magnetic  susceptibility, \Tkai{}, along the three principal axes of Sr$_2$Cu$_{1-x}$Pd$_x$O$_3$ with $x = 0$, $0.004$, $0.010$. 
It is found that \Tkai{} is almost constant at high temperatures and increases with decreasing temperature at low temperatures below $50~\TK$. 
The \Tkai{} is so anisotropic that the value of \Tkai{} along the $c$-axis is larger than those along the $a$- and $b$-axes. 
The increase of \Tkai{} at low temperatures is Curie-like and becomes marked with increasing $x$. 
These behaviors of \Tkai{} are similar to those reported in the literature \cite{Ami:PRB51:1995:5994,Motoyama:PRL76:1996:3212,Kojima:PRB70:2004:094402}.

The experimental data of \Tkai{} are well fitted using the following equation as in the former report \cite{Motoyama:PRL76:1996:3212}: 
\begin{equation}
	\Tkai{} = \Tkai{Curie} + \Tkai{spin} + \Tkai{0} ,
	\label{equ:KaiAll}
\end{equation}
where \Tkai{Curie} is the Curie term, \Tkai{spin} is the contribution of spin chains and \Tkai{0} is a constant term due to the Van Vleck paramagnetism and the ion-core diamagnetism. 
The \Tkai{Curie} is given by
\begin{equation}
	\Tkai{Curie}
		= \frac{\Tsub{N}{A}g^2\Tsub{\mu}{B}^2S(S+1)\Tsub{x}{Curie}}
		       {3\TkB(T-\theta)}, 
	\label{equ:KaiCurie}
\end{equation}
where \Tsub{N}{A} is Avogadro's number, $g$ the g-factor, \Tsub{\mu}{B} the Bohr magneton, \Tsub{x}{Curie} the concentration of free spins per Cu, {\TkB} the Boltzmann constant and $\theta$ the Weiss temperature. 
The \Tkai{spin} is given by the following equation proposed by Eggert \textit{et al}. \cite{Eggert:PRL73:1994:332} for the 1D $S=1/2$ Heisenberg AF spin system at low temperatures of $\TkB T \ll \TJ{}{}$: 
\begin{equation}
	\Tkai{spin}
		= \frac{1}{\pi ^2 \TJ{}{}} 
		  \left( 1 + \frac{1}{2 \ln \left( \TT{0} / T \right) } \right), 
	\label{equ:KaiSpin}
\end{equation}
where \TJ{}{} is the nearest neighbor exchange interaction and 
\TT{0} is a parameter depending on the second nearest neighbor exchange interaction. In the fitting we assume that $\TT{0} = 7.7 \TJ{}{}$ which is true in the case that the second nearest neighbor exchange interaction is neglected. 
The value of the ion-core diamagnetism is put at $-10.7 \times 10^{-5}$ emu/mol \cite{TEMS:1932}. 
The value of the Van Vleck paramagnetism is fixed at $4.7 \times 10^{-5}$ emu/mol for the $a$- and $b$-axes and $9.7 \times 10^{-5}$ emu/mol for the $c$-axis, because its accurate value is not obtained from the fitting. 
Although the fixation of the \Tkai{0} value in the fitting enhances the uncertainty of the obtained value of \TJ{}{}, it is noted that the change of \Tkai{0} and \TJ{}{} values does not affect the value of the Curie term so much, because the Curie term is sensitive to only the increase of the magnetic susceptibility at low temperatures.

\begin{table}[tbp]
	\caption{Parameters used for the fit of the temperature dependence of the magnetic susceptibility, \Tkai{}, in Sr$_2$Cu$_{1-x}$Pd$_x$O$_3$ with Eqs. \Teqref{equ:KaiAll} -- \Teqref{equ:KaiSpin}.}
	\label{tab:sup}
	\begin{center}
	\begin{tabular}{cc|*{3}{c}}
	\hline
	\hline
		\multicolumn{1}{c} {\TTabSpace{$x$}} &
		\multicolumn{1}{c|}{\TTabSpace{axis}} &
		\multicolumn{1}{c} {\TTabSpace{\Tsub{x}{Curie}}} &
		\multicolumn{1}{c} {\TTabSpace{$\theta~({\TK})$}} &
		\multicolumn{1}{c} {\TTabSpace{$\TJ{}{}~({\TK})$}} \\
	\hline
				& $a$	& 0.0013(1) & -3.54(14) & 1680(7) \\
		0		& $b$	& 0.0011(1) & -2.99(4) & 1780(2) \\
				& $c$	& 0.0007(1) & -1.97(13) & 1800(7) \\
	\hline
				& $a$	& 0.0024(1) & -0.35(12) & 1650(44) \\
		0.004	& $b$	& 0.0028(1) & -0.71(9) & 1720(29) \\
				& $c$	& 0.0027(1) & -0.65(7) & 2010(37) \\
	\hline
				& $a$	& 0.0040(1) & -0.88(6) & 1980(40) \\
		0.010	& $b$	& 0.0039(1) & -0.74(6) & 2230(51) \\
				& $c$	& 0.0046(1) & -0.77(6) & 2220(62) \\
	\hline
	\hline
	\end{tabular}
	\end{center}
\end{table}

The best-fit results are shown by solid lines in Fig. \ref{fig:sup}.
Parameters obtained from the best fit are listed in Table \ref{tab:sup}. 
The value of \TJ{}{} is estimated to be $1900 \pm 300~\TK$. 
This value is slightly smaller than that in the former report \cite{Motoyama:PRL76:1996:3212}, which may be due to the difference of the fitting temperature-range. 
Values of $\theta$ show no systematic change in correspondence with the \Tion{Pd}{2+} doping, as in the former report \cite{Kojima:PRB70:2004:094402}. 
The value of \Tsub{x}{Curie} is found to be about a half of the Pd concentration, $x$. 
This is reasonably explained as follows. 
That is, in an 1D AF spin system, the Curie term originates from free spins due to finite spin segments separated by spin defects such as \Tion{Pd}{2+}, lattice defects and unintended impurities \cite{COMMENT:Sr213Imp}. 
Therefore, there exist finite spin segments with a free spin of $S = 1/2$ composed of an odd number of spins (called odd spin segments) and those with $S = 0$ composed of an even number of spins (called even spin segments).
An even spin segment is divided by one spin-defect into an even spin segment and an odd spin segment. 
On the other hand, an odd spin segment is divided by one spin-defect into two even segments or two odd segments.
Therefore, one free spin of $S = 1/2$ is induced, on average, by two spin-defects \cite{Eggert:PRL89:2002:047202}. 
Accordingly, it is reasonable that the value of \Tsub{x}{Curie} is about a half of $x$. 
The average length of finite spin chains, namely, \Tsub{L}{imp} can be calculated from the following equation:
\begin{equation}
	\Tsub{L}{imp} =\frac{c}{2 \Tsub{x}{Curie}},
	\label{equ:kaiLimp}
\end{equation}
where \Tsub{x}{Curie} is the average value of \Tsub{x}{Curie}'s in each $x$ and $c$ is the lattice parameter along the $c$-axis; $c = 3.499$ {\AA}.
Values of \Tsub{L}{imp} are listed in Table \ref{tab:lsp}, using the average value of \Tsub{x}{Curie}'s obtained from the magnetic susceptibility measurements in fields parallel to the three principal axes.

\subsection{Specific heat}
\begin{figure}[tbp]
	\begin{center}
		\includegraphics[scale=\TFigScale]{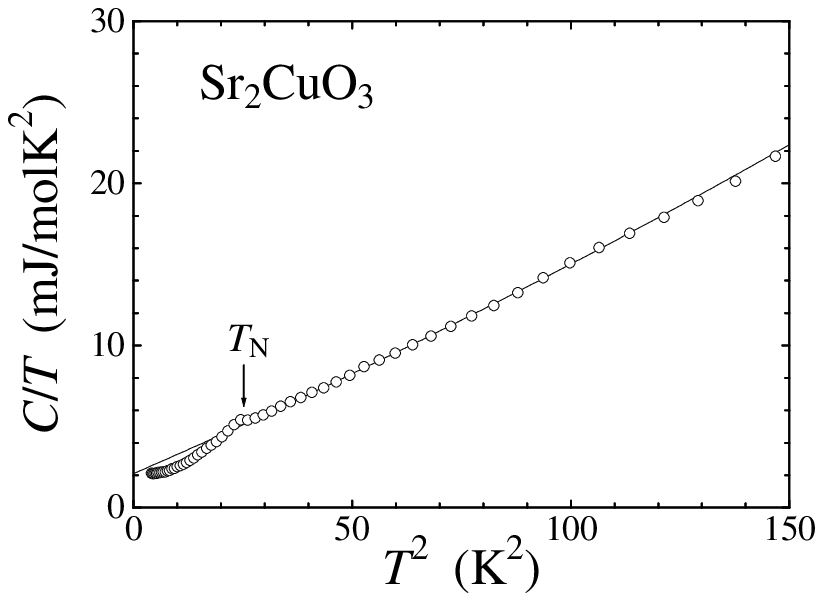}
		\caption{Temperature dependence of the specific heat, \TC{}, of Sr$_2$CuO$_3$. 
		The solid line is the best-fit result using Eqs. \Teqref{equ:cap} -- \Teqref{equ:capPhonon}. }
		\label{fig:cap}
	\end{center}
\end{figure}
Figure \ref{fig:cap} shows the temperature dependence of the specific heat, $C$, of Sr$_2$CuO$_3$. 
It is found that $C$ exhibits a small peak around $5$ K 
which is in good agreement with \TT{N} observed in the neutron scattering \cite{Kojima:PRL78:1997:1787} and {\TmSR} measurements \cite{Keren:PRB48:1993:12926,Kojima:PRL78:1997:1787}. 
Since there is no electronic contribution to $C$ in Sr$_2$CuO$_3$, $C$ is given by the sum of the magnetic specific heat due to spin chains, namely, \TC{spinon} and the specific heat of phonons, \TC{phonon}, as the follows: 
\begin{equation}
	C = \TC{spinon} + \TC{phonon}.
	\label{equ:cap}
\end{equation}
The \TC{spinon} is given by the following equation based on the 1D AF Heisenberg model with $S = 1/2$ at low temperatures of $\TkB T \ll \TJ{}{}$ \cite{Takahashi:PTP50:1973:1519}: 
\begin{equation}
	\TC{spinon} = \frac{2 \Tsub{N}{s} \TkB ^2}{3 J} T, 
	\label{equ:capSpin}
\end{equation}
where \Tsub{N}{s} is the number of spins. 
The \TC{phonon} is given by the following equation based on the Debye model: 
\begin{equation}
	\TC{phonon} = \frac{12 \pi ^4 N \TkB}{5 \TDebye ^3} T^3 + \delta T^5, 
	\label{equ:capPhonon}
\end{equation}
where $N$ is the number of atoms and $\delta$ is the coefficient of the anharmonic term. 
The data of \TC{} at low temperatures above \TT{N} are well fitted using Eqs. \Teqref{equ:cap} -- \Teqref{equ:capPhonon}, as shown by the solid line in Fig. \ref{fig:cap}.
Values of the best-fit parameters are $\TJ{}{} = 2634 \pm 150~\TK$, $\TDebye = 470.8 \pm 8.2~\TK $ and $ \delta = 1.2 \pm 0.2 \times 10^{-7}$~J/K$^6$ mol.
These values are comparable with those in the former report, respecively \cite{Sologubenko:PRB62:2000:R6108}. 
However, the value of \TJ{}{} is a little larger than that estimated from magnetic susceptibility measurements, which may be due to the difference of the fitting temperature-range bewteen the two kinds of measurement. 

\subsection{Thermal conductivity}
\begin{figure*}[tbp]
	\begin{center}
		\includegraphics[scale=\TFigScaleL]{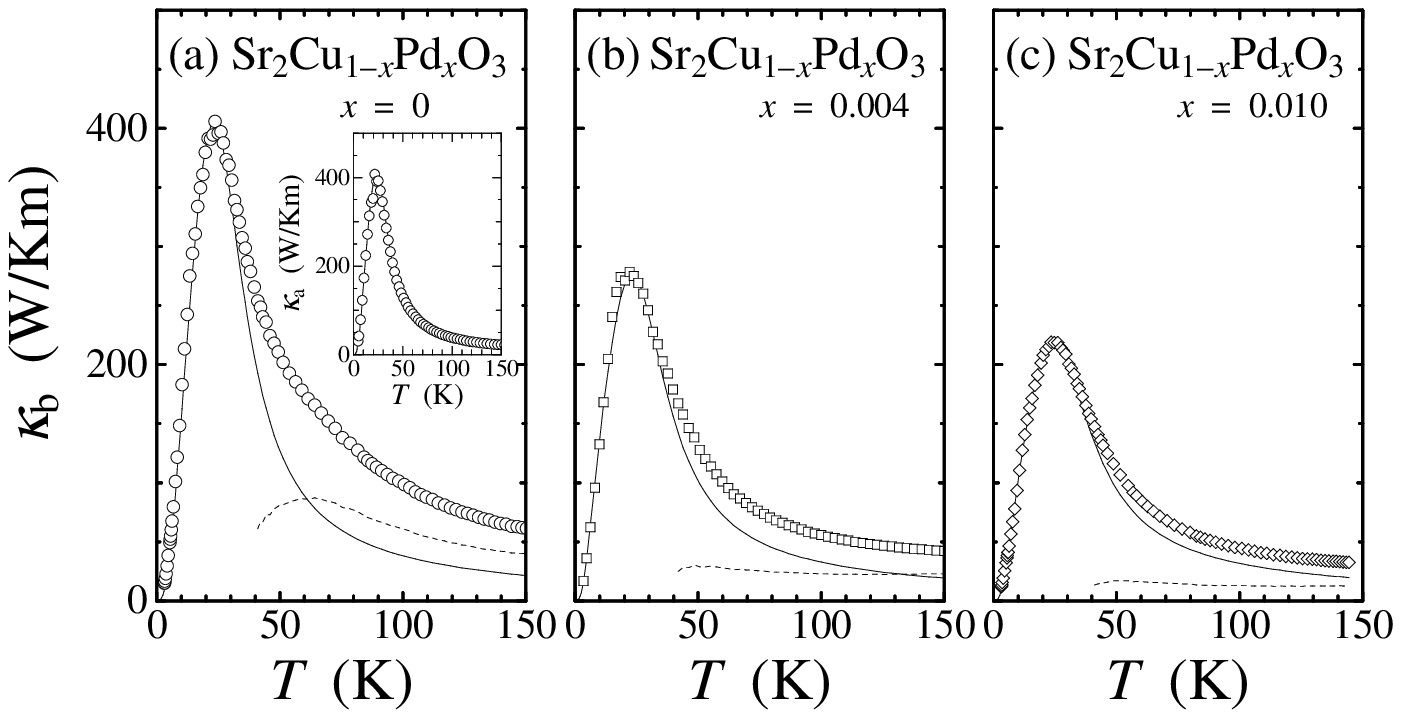}
		\caption{Temperature dependence of the thermal conductivity along the $b$-axis parallel to the spin chain, \Tk{b}, of Sr$_2$Cu$_{1-x}$Pd$_x$O$_3$ with (a) $x = 0$, (b) $0.004$, (c) $0.010$. 
		The inset shows the temperature dependence of the thermal conductivity along the $a$-axis perpendicular to the spin chain, \Tk{a}, of Sr$_2$CuO$_3$. 
		Solid lines are \Tk{phonon} estimated using Eqs. \Teqref{equ:kph} -- \Teqref{equ:tauph} on the Debye model. 
		Dashed lines are \Tk{spinon} obtained by subtracting \Tk{phonon} from \Tk{b}.}
		\label{fig:kap}
	\end{center}
\end{figure*}
Figure \ref{fig:kap} shows the temperature dependence of \Tk{b} along the spin chain of Sr$_2$Cu$_{1-x}$Pd$_x$O$_3$ with (a) $x = 0$, (b) $x = 0.004$ and (c) $x = 0.010$. 
The temperature dependence of \Tk{a} perpendicular to the spin chain in $x = 0$ is also shown in the inset of Fig. \ref{fig:kap} (a). 
For $x = 0$, it is found that \Tk{a} increases with decreasing temperature and exhibits a peak around $25$ K. 
This is a typical behavior of the thermal conductivity due to phonons, \Tk{phonon}. 
On the other hand, \Tk{b} increases with decreasing temperature and exhibits a small shoulder due to \Tk{spinon} around $70$ K in addition to the peak due to \Tk{phonon} around $25$ K. 
These behaviors are similar to those in the former report \cite{Sologubenko:PRB62:2000:R6108,Sologubenko:PRB64:2001:054412}. 
For $x$ = 0.004, 0.010, both the peak around $25$ K and the small shoulder around $70$ K are suppressed by the \Tion{Pd}{2+} doping. 
These results indicate that both phonons and spinons are scattered by nonmagnetic impurities of \Tion{Pd}{2+} so that their mean free paths become short.

Here, we estimate \Tk{spinon}. 
For this purpose, at first, the estimate of \Tk{phonon} is necessary. 
The \Tk{phonon} is given by the following equation based on the Debye model \cite{TCS:1976}.
\begin{equation}
	\Tk{phonon} = \frac{ \TkB }{2 \pi^2 \Tv{phonon}}
	              \left( \frac{ \TkB T }{ \hbar } \right) ^3
	              \int _0 ^{ \TDebye /T }
	              \frac{x^4 \mathrm{e}^x}{(\mathrm{e}^x - 1) ^2}
	              \Tt{phonon}
	              \mathrm{d} x, 
	\label{equ:kph}
\end{equation}
where $x = \hbar \omega / \TkB T $, $\omega$ is the phonon angular frequency, $\hbar$ the Planck constant, \Tv{phonon} the phonon velocity and $\Tt{phonon}$ the relaxation time of the phonon scattering. 
The \Tv{phonon} is calculated as 
\begin{equation}
	\Tv{phonon} = \frac{ \TkB \TDebye }{ \hbar} (6 \pi ^2 n) ^{-1/3}, 
	\label{equ:vph}
\end{equation}
where $n$ is the number density of atoms. 
The phonon scattering rate, $\Tt{phonon}^{-1}$, is assumed to be given by the sum of scattering rates due to various scattering processes as follows, 
\begin{equation}
	\Tt{phonon} ^{-1}
	=
	\frac{ \Tv{phonon} }{ \Tsub{L}{b} }
	+
	A \omega ^4
	+
	B \omega ^2 T \exp \left( - \frac{ \TDebye }{ bT } \right)
	, 
	\label{equ:tauph}
\end{equation}
where \Tsub{L}{b}, $A$, $B$ and $b$ are fitting parameters. 
The first term represents the phonon scattering by boundaries; the second, the phonon scattering by point defects; the third, the phonon-phonon scattering in the umklapp process. 
Using Eqs. \Teqref{equ:kph} -- \Teqref{equ:tauph} and putting {\TDebye} at $470.8$ K from the specific heat measurements, the data of \Tk{a} in $x = 0$ are well fitted, as shown by the solid line in the inset of Fig. 3 (a). 
The estimated \Tk{phonon} in \Tk{b} is performed by the fit of the data of \Tk{b} at low temperatures below 25 K with Eqs. \Teqref{equ:kph} -- \Teqref{equ:tauph}. 
In the fitting, {\TDebye} is put at 470.8 K. 
Values of $B$ and $b$ are put at the same values as these used for the fit of \Tk{a} in $x = 0$, respectively, because the phonon-phonon scattering in the umklapp process seems neither to be affected by the direction nor by the slight doping of \Tion{Pd}{2+} so much. 
The adjusting parameters are only \Tsub{L}{b} and $A$, which depend on the phonon scattering by boundaries and by point defects, respectively. 
Then, \Tk{spinon} is estimated by subtracting the fitting curve of \Tk{phonon} from the data of \Tk{b}, as shown by dashed lines in Fig. \ref{fig:kap}. 
Here, it is noted that \Tk{spinon} is a little underestimated, because \Tk{spinon} is neglected at low temperatures below 25 K. 
Values of the parameters used for the best fit are listed in Table \ref{tab:kap}. 
These are comparable with those in the former report \cite{Sologubenko:PRB62:2000:R6108,Sologubenko:PRL84:2000:2714,Sologubenko:PRB64:2001:054412,Sologubenko:EL62:2003:540}. 
It is found that the value of $A$ increases with increasing $x$. 
This is resonable, because phonons are scattered by substituted \Tion{Pd}{2+} ions. 

\begin{table}[tbp]
	\caption{Parameters used for the fit of the temperature dependence of the thermal conductivity, \Tk{}, in Sr$_2$Cu$_{1-x}$Pd$_x$O$_3$ with Eqs. \Teqref{equ:kph} -- \Teqref{equ:tauph}.}
	\label{tab:kap}
	\begin{center}
	\begin{tabular}{cc|cccc}
	\hline
	\hline
		\multicolumn{1}{c} { \TTabSpace{$x$} }	&
		\multicolumn{1}{c|}{ \TTabSpace{axis} }	&
		\multicolumn{1}{c} { \TTabSpace{\Tsub{L}{b} ({m})} }	&
		\multicolumn{1}{c} { \TTabSpace{$A$ (s$^3$)} }	&
		\multicolumn{1}{c} { \TTabSpace{$B$ (s/K)} }	&
		\multicolumn{1}{c} { \TTabSpace{$b$} }	\\
	\hline
		0		&	$a$ &	$6.00 \times 10^{-4}$ &	$1.38 \times 10^{-44}$ &	$2.70 \times 10^{-18}$ &	3.49 \\
				&	$b$ &	$6.00 \times 10^{-4}$ &	$1.38 \times 10^{-44}$ &	$2.70 \times 10^{-18}$ &	3.49 \\
	\hline
		0.004	&	$b$ &	$9.00 \times 10^{-4}$ &	$3.96 \times 10^{-44}$ &	$2.70 \times 10^{-18}$ &	3.49 \\
	\hline
		0.010	&	$b$ &	$5.90 \times 10^{-4}$ &	$4.57 \times 10^{-44}$ &	$2.70 \times 10^{-18}$ &	3.49 \\
	\hline
	\hline
\end{tabular}
\end{center}
\end{table}

\begin{figure}[tbp]
	\begin{center}
		\includegraphics[scale=\TFigScale]{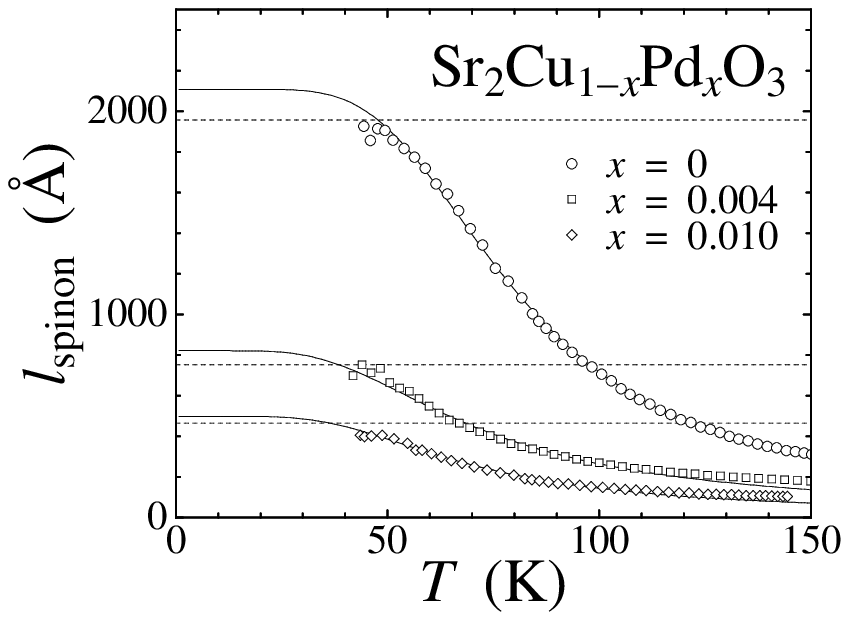}
		\caption{Temperature dependence of the mean free path of spinons, \Tl{spinon}, of Sr$_2$Cu$_{1-x}$Pd$_x$O$_3$ with $x = 0$, $0.004$, $0.010$. 
		Solid lines are the best-fit results using Eq. \Teqref{equ:lspfit}. 
		Dashed lines are the average length between spin defects, \Tsub{L}{imp}, estimated from the magnetic susceptibility measurements for $x$ = 0 (upper), 0.004 (middle), 0.010 (lower).}
		\label{fig:lsp}
	\end{center}
\end{figure}

Next, we estimate \Tl{spinon} using the following equation, 
\begin{equation}
	\Tk{spinon} = \TC{spinon} \Tv{spinon} \Tl{spinon}, 
	\label{equ:ksp}
\end{equation}
where \Tv{spinon} is the velocitiy of spinons. 
The \Tv{spinon} is given by the following equation based on the des Cloizeaux-Pearson mode at low temperatures of $\TkB T \ll J$ \cite{Cloizeaux:PR128:1962:2131}: 
\begin{equation}
	\Tv{spinon} = \frac{\pi J a}{2 \hbar}, 
	\label{equ:vsp}
\end{equation}
where $a$ is the  distance between the nearest neighboring spins in the chain. 
Therefore, \Tl{spinon} is calculated using Eq. \Teqref{equ:capSpin} as follows, 
\begin{equation}
	\Tl{spinon} = \frac{3 \hbar}{\pi \Tsub{N}{s} a \TkB ^2 T} \Tk{spinon}. 
	\label{equ:lsp}
\end{equation}
Figure \ref{fig:lsp} shows the temperature dependence of \Tl{spinon} obtained thus and \Tsub{L}{imp} estimated from the magnetic susceptibility measurements. 
It is found that \Tl{spinon} increases with decreasing temperature and seems to be saturated at low temperatures. 
Since \Tk{spinon} is neglected at low temperatures below 25 K in this analysis as mentioned above, values of \Tl{spinon} at low temperatures are uncertain. 
Therefore, values of \Tl{spinon} at low temperatures are estimated as shown by solid lines in Fig. \ref{fig:lsp}, by fitting the data of \Tl{spinon} at high temperatures above 55 K with the following simple equation \cite{Sologubenko:PRB62:2000:R6108,Sologubenko:PRB64:2001:054412,Sologubenko:EL62:2003:540}: 
\begin{eqnarray}
	\Tl{spinon} ^{-1} = \Tsub{A}{s} T \exp (- T ^* / T) + L^{-1},
	\label{equ:lspfit}
\end{eqnarray}
where \Tsub{A}{s}, {\TTst} and $L$ are fitting parameters. 
The first term is due to the spinon-phonon and/or spinon-spinon scattering in the umklapp process and {\TTst} is the characteristic temperature. 
The second term is due to the spinon scattering by spin defects. 
The value of $L$ correspounds to the saturated value of \Tl{spinon} at low temperatures. 
Parameters obtained from the best fit are listed in Table \ref{tab:lsp}. 
It is found that the value of \Tsub{A}{s} inreases with increasing $x$. 
Since the \Tion{Pd}{2+} doping is guessed to induce local phonons scattering spinons around \Tion{Pd}{2+}, the spinon-phonon interaction may increase with increasing $x$.
To our surprise in Table \ref{tab:lsp}, values of $L$ decrease with increasing $x$ and are very close to those of \Tsub{L}{imp}, respectively. 
This means that \Tl{spinon} at low temperatures is approximately limited by \Tsub{L}{imp}. 
That is, the thermal conduction due to spinons is limited at low temperatures only by scattering by spin defects. 
Accordingly, it is concluded that the thermal conduction due to spinons at low temperatures is ballistic as theoretically expected \cite{Castella:PRL74:1995:972,Saito:PRE54:1996:2404,Zotos:PRB55:1997:11029,Zotos:PRL82:1999:1764,Klumper:JPA35:2002:2173}.

\begin{table}[tbp]
	\caption{Parameters used for the fitting of the temperature dependence of the mean free path of spinons, \Tl{spinon}, with Eq. \Teqref{equ:lspfit} in Sr$_2$Cu$_{1-x}$Pd$_x$O$_3$. 
	The average value of the concentration of free spins per Cu, \Tsub{x}{Curie}, and the average length between spin defects, \Tsub{L}{imp}, estimated from the magnetic susceptibility measurements are also listed. }
	\label{tab:lsp}
	\begin{center}
	\begin{tabular}{ccrrcc}
	\hline
	\hline
		\multicolumn{1}{c}{ \TTabSpace{$x$} }				&
		\multicolumn{1}{c}{ \TTabSpace{\Tsub{x}{Curie}} }	&
		\multicolumn{1}{c}{ \TTabSpace{\Tsub{L}{imp} (\AA)} }&
		\multicolumn{1}{c}{ \TTabSpace{$L$~(\AA)} }			&
		\multicolumn{1}{c}{ \TTabSpace{\Tsub{A}{s}~(s/K)} }	&
		\multicolumn{1}{c}{ \TTabSpace{\Tsup{T}{*}~(K)} }	\\
	\hline
		0		& $0.0010$	& $1960$ \hspace{1em}	& $2110 \pm 26$ \hspace{0em}	& $8.89 \pm 0.47 \times 10^{-5}$	& $227 \pm 5$ \\
	\hline
		0.004	& $0.0026$	& $753$ \hspace{1em}	& $821 \pm 83$ \hspace{0em}	& $10.1 \pm 1.8 \times 10^{-5}$	& $136 \pm 22$ \\
	\hline
		0.010	& $0.0042$	& $466$ \hspace{1em}	& $498 \pm 60$ \hspace{0em}	& $20.9 \pm 4.2 \times 10^{-5}$	& $145 \pm 24$ \\
	\hline
	\hline
\end{tabular}
\end{center}
\end{table}

\section{Conclusion}
We have measured the thermal conductivity, magnetic susceptibility and specific heat of Sr$_2$Cu$_{1-x}$Pd$_x$O$_3$ single crystals with $x = 0$, $0.004$ and $0.010$ which are regarded as an 1D $S=1/2$ Heisenberg AF spin system described by an integrable Hamiltonian, in order to prove the theoretical prediction that the thermal conduction due to spinons is ballistic. 
In this system, the length of the spin chain between spin defects is controlled by the doping of nonmagnetic impurities of \Tion{Pd}{2+}. 
We have estimated the average length of finite spin chains between spin defects, \Tsub{L}{imp}, from the magnetic susceptibility measurements and the mean free path of spinons, \Tl{spinon}, from the thermal conductivity measurements using the Debye temperature estimated from the specific heat measurement. 
It has been found that values of \Tl{spinon} at low temperatures for $x = 0$, $0.004$ and $0.010$ are very close to those of \Tsub{L}{imp}, respectively. 
This means that spinons carry heat along the spin chain between spin defects without being scattered at low temperatures. 
Accordingly, our results strongly support the theoretical prediction that the thermal conduction due to spinons in 1D $S=1/2$ spin systems described by intgrable Hamiltonian's is ballistic. 

\section*{Acknowledgments}
We are grateful to Prof. S. Maekawa, Prof. T. Tohyama and Dr. H. Tsuchiura for the helpful discussion. 
The thermal conductivity measurements were performed at the High Field Laboratory for Superconducting Materials, Institute for Materials Research, Tohoku University. This work was partly supported by a Grant-in-Aid for Scientific Research from the Ministry of Education, Culture, Sports, Science and Technology, Japan. One of the authors (T. K.) was supported by the Japan Society for the Promotion of Science.


\begin{thebibliography}{10}
\bibitem{COMMENT:Sr213Imp}
Since the purity of the raw material of CuO was 99.9 \%, the grown crystals of
  Sr$_2$Cu$_{1-x}$Pd$_x$O$_3$ should include unknown impurities less than 0.1
  \%. This is a reason why $x_{\mathrm{Curie}}$ is not zero for the Pd-free
  crystal of $x=0$.
\end{thebibliography}
\end{document}